\documentclass[twocolumn,showpacs,preprintnumbers,amsmath,amssymb]{revtex4}

\usepackage{graphicx}
\usepackage{dcolumn}
\usepackage{bm}

\begin{document}


\title{Chaos and its quantization in dynamical Jahn-Teller systems}

\author{Hisatsugu  Yamasaki}
\email{hisa@physics.s.chiba-u.ac.jp}
\homepage{http://zeong.s.chiba-u.ac.jp/~hisa/}
\author{Yuhei Natsume}%
\affiliation{%
Graduate School of Science and Technology, Chiba-University\\
Inage-ku, Chiba, 263-8522 Japan
}%

\author{Akira Terai}
\author{Katsuhiro Nakamura}
\affiliation{
Department of Applied Physics,
Osaka City University\\
Sumiyoshi-ku
Osaka, 558-8585 Japan
}%
     
\date{\today}

\begin{abstract}
We investigate the $E_g \otimes e_g$ Jahn-Teller system for the
purpose to reveal the nature of quantum chaos in crystals.
This system simulates the
interaction between the nuclear vibrational modes and
the electronic motion in non-Kramers doublets for multiplets of
transition-metal ions.
Inclusion of the anharmonic potential due to the trigonal symmetry in crystals
makes the system nonintegrable and chaotic. Besides the quantal analysis of the transition from Poisson to Wigner level statistics with increasing
the strength of anharmonicity, we study the effect of chaos on
the electronic orbital angular momentum and explore the
magnetic $g$-factor as a function of the system's energy. The regular
 oscillation of this factor changes to a rapidly-decaying
irregular oscillation by increasing the anharmonicity (chaoticity).

\end{abstract}

\pacs{05.45.Mt,71.70.Ej,82.90.+j,03.65.-w,31.30.Gs}
\maketitle

\section{Introduction}
Recently the study on quantization of classically chaotic Hamiltonian
systems has received a wide attention. An accumulation of numerical and
experimental data indicates  Wigner-type level statistics, wavefunction
scars and other
characteristic features\cite{9,11}.

In addition to toy models like a kicked rotator, H{\'e}non-Heiles 
system, etc.,
some realistic systems like a
hydrogen atom in a magnetic field and micro-wave cavities are also being
investigated\cite{9,11,27}. Quantum mechanics of chaotic systems also suggests insight
beyond a simple quantal manifestation of chaos\cite{28,29}. Therefore it is crucial to 
have more
and more experimentally-accessible quantum systems which exhibit chaos in its
classical treatment.

In this paper we choose the Jahn-Teller system simulating
transition-metal ions embedded in the host crystals such as III-V
semiconductors and halides crystals. Among them we consider $E_g \otimes 
e_g$ model
associated with the irreducible representation for the cubic symmetry group,
namely, the two-dimensional (2-d)
lattice-vibration modes $e_g$
linearly coupled to doubly-degenerate electronic states $E_g$\cite{4}. This
system has an adiabatic doubly-fold lattice potential with
the conical intersection
of the potential surfaces, whose geometric phase was one of topics some
time ago\cite{1,2}. The lattice potential here can be harmonic or anharmonic.
 From the classical dynamical viewpoint in the adiabatic limit, as shown below, the system
with the 2-d harmonic potential is
integrable, leading to regular motions, and on adding the anharmonic
term, it becomes nonintegrable and chaotic\cite{3}.
A systematic investigation of the
quantal counterpart of classical chaos in these
systems is
desirable. Furthermore, since the model is a representative for
paramagnetic ions, it is experimentally important to see the effect of
chaos on the magnetic $g$-factor.
This factor is an expectation value for electronic orbital angular momentum
and measures a degree of level splitting of highly excited states
induced by the lattice-electron interaction. The
oscillating structure in energy dependence of $g$-factor
is expected to reflect
the feature of the underlying classical dynamics.

The organization of the paper is as follows: In Section II a model
for  Jahn-Teller  $E_g \otimes e_g$ system is proposed.  Section III deals 
with the
classical analyses of the model. Both the systems with and without
anharmonic terms are examined.  Section IV presents a quantization of the
system together with level statistics. Section V is concerned with a
proposal of the experiment to verify the quantum signature of
chaos in the dynamical Jahn-Teller system.
Some novel feature of $g$-factor is explored there.  The
final Section is devoted to summary and discussions.

\section{\label{sec:level1}Dynamical Jahn-Teller System}

\begin{figure}[htbp]
  \begin{minipage}{.47\textwidth}
\includegraphics[width=.80\linewidth]{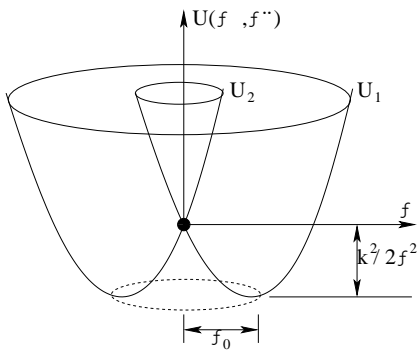}
\caption{Adiabatic potential of
Mexican-hut shape. $\rho =\sqrt{Q_1^2+Q_2^2}$.
The potential has the degenerate minimum.}
   \label{fig1}
  \end{minipage}
\end{figure}

We investigate the electronic states of degenerate $E_g$ orbitals
of $d$-levels in transition-metal ions coupled with 2-d vibrational modes
$e_g$ expressed by coordinates $Q_1$ and $Q_2$.
The $E_g \otimes e_g$ model is the typical system showing dynamic Jahn-Teller effects (DJTE),
which has been discussed
in the field of magnetism
for transition-metal ions\cite{4,5}.
The Hamiltonian matrix $H$
for this system is expressed as
\begin{eqnarray}
H &=& -\frac{\hbar^2}{2}
(\frac{\partial^2}{\partial Q_1^2}+\frac{\partial^2}{\partial Q_2^2}){\bf I}
+k
\left[
\begin{array}{cc}
  Q_1  & Q_2  \\
  Q_2  &  -Q_1
\end{array}
\right] \nonumber \\
&+& V({\bf Q}){\bf I},
\label{eqn:1}
\end{eqnarray}
where {\bf I} is the $2\times 2$ unit matrix and $V({\bf Q})$ is a potential 
energy.
The nuclear mass is set to be unity. The second term of (\ref{eqn:1}) is the so-called Jahn-Teller interaction, $H_{J-T}$ with
$k$ the coupling parameter between
electronic states and vibrational modes.
Bases for electronic orbitals $E_g$ lying behind (\ref{eqn:1}) are
$\langle{\bf r}|u\rangle=u({\bf r})=3z^2-r^2$ and $\langle{\bf r}|v\rangle=v({\bf r})=x^2-y^2$.
When $V({\bf Q})$ is a harmonic potential given by
\begin{equation}
V_0({\bf Q})=\frac{1}{2}\omega^2(Q_1^2+Q_2^2), \label{eqn:2}
\end{equation}
the corresponding adiabatic potential for (\ref{eqn:1}) has
the Mexican-hut shape in Fig.\ref{fig1},
where $Q_1=\rho\cos\theta$ and $Q_2=\rho\sin\theta$.
The potential minima lie at
$\rho=\rho_0=k/\omega^2$ with an arbitrary value of $\theta$.
Namely the minima are infinitely degenerate.
The energy for the minima is $k^2/2\omega^2$.
Vibronic levels for the quantum Hamiltonian (\ref{eqn:1})
were discussed in numerical calculations using
small dimensional Hamiltonian matrices\cite{7}.
Recently this model has been investigated from a viewpoint of the
geometric phase\cite{5,6}.
On the other hand, the effect of the trigonal fields
expressed as the anharmonic term
\begin{equation}
V_A({\bf Q})=-(b/3)(Q_1^3 -3 Q_1 Q_2^2)\label{eqn:3}
\end{equation}
was also analysed
as to some low-lying levels \cite{7,8}.
In short, O'Brien investigated the system (\ref{eqn:1}) with
potential $V({\bf Q})=V_0({\bf Q})+V_A({\bf Q})$
in the low-energy approximation
that $\rho$ is fixed to $\rho_0$.
However, we numerically calculate eigenvalues and eigenvectors without having recourse to
such an approximation.
We derive level spacing distributions to see the effect of chaos on quantum
systems\cite{9,11,13,14,15,27}.
Furthermore we investigate the quantal
$g$-factor, whose oscillating structure was shown three decades ago
by Washimiya in the system
without anharmonicity. We explore the effect of chaos on the $g$-factor in
the system with the anharmonicity. The dynamical Jahn-Teller system was also
studied by Bulgac and
Kusnezov\cite{20,21,22,23}
in a system with the three-dimensional harmonic potential.
However, we should note that
the dimensionality of lattice-vibration modes
characterized by the irreducible representation is two and not three
according to the theory of a point-symmetry group applied to real crystals
and that our model is a better reflection of the real
crystal\cite{4}.

\section{Quasi-Classical dynamics and chaos}
In the first place, we shall analyze the quasi-classical counterpart
of Hamiltonian (\ref{eqn:1}), which is given by
\begin{equation}
  H=\frac{1}{2}(P_1^2+P_2^2)+V({\bf Q})+k({\bf Q}\cdot{\bf \sigma}),
\label{eqn:4}
\end{equation}
where the first term is a kinetic energy for classical vibrational modes with
coordinates ${\bf Q}=(Q_1,Q_2)$ and the second one is the harmonic and/or anharmonic potential;
the third one
is the quasi-classical form for the Jahn-Teller interaction where
${\bf \sigma}$ is Pauli matrices
$\sigma=(\sigma_x,\sigma_y,\sigma_z)$. Noting that ${\bf \sigma}$ space is independent of the real space, we choose $\sigma_x,\sigma_y$ and $\sigma_z$ corresponding to $\sigma_2,\sigma_3$ and $\sigma_1$, respectively.
It is convenient to represent the quantum state with use of the
density matrix ${\bf \rho}$
\begin{equation}
  {\bf \rho}=\frac{1}{2}
\left(
\begin{array}{cc}
  1+z  & x-iy \\
  x+iy & 1-z
\end{array}
\right),\label{eqn:5}
\end{equation}
where ${\bf r}={\bf Tr}({\bf \rho \sigma})=(x,y,z)\equiv ({\bf 
r_{\perp}},z)$ is a
real vector. Using the potential
\begin{equation}
V({\bf Q})=V_0({\bf Q})+V_A({\bf Q})
  \label{eqn:6}
\end{equation}
with $V_0$ and $V_A$ in (\ref{eqn:2}) and (\ref{eqn:3}), the
equations
of motion derived from (\ref{eqn:4}) are
\begin{subequations}
\begin{align}
\frac{d{\bf Q}}{dt}&={\bf P} \label{eqn:7_a} \\ 
\frac{d{\bf P}}{dt}&=-\frac{dV({\bf Q})}{d{\bf Q}}-k{\bf r_{\perp}} \label{eqn:7_b}\\
\frac{d{\bf r}}{dt}&=k{\bf Q}\times{\bf r}.\label{eqn:7_c}
\end{align}
\end{subequations}
\begin{figure*}[htbp]
\begin{minipage}{.47\textwidth}
\includegraphics[width=\linewidth]{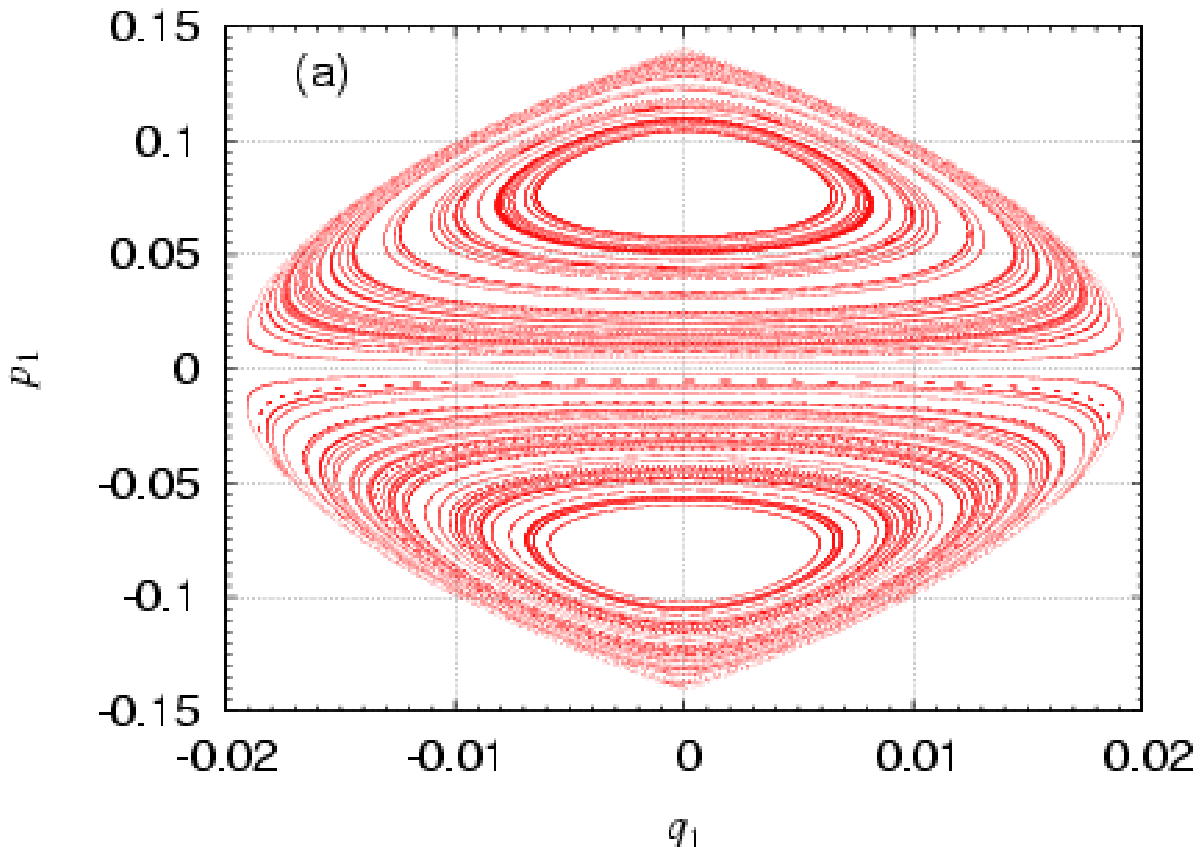}
\includegraphics[width=\linewidth]{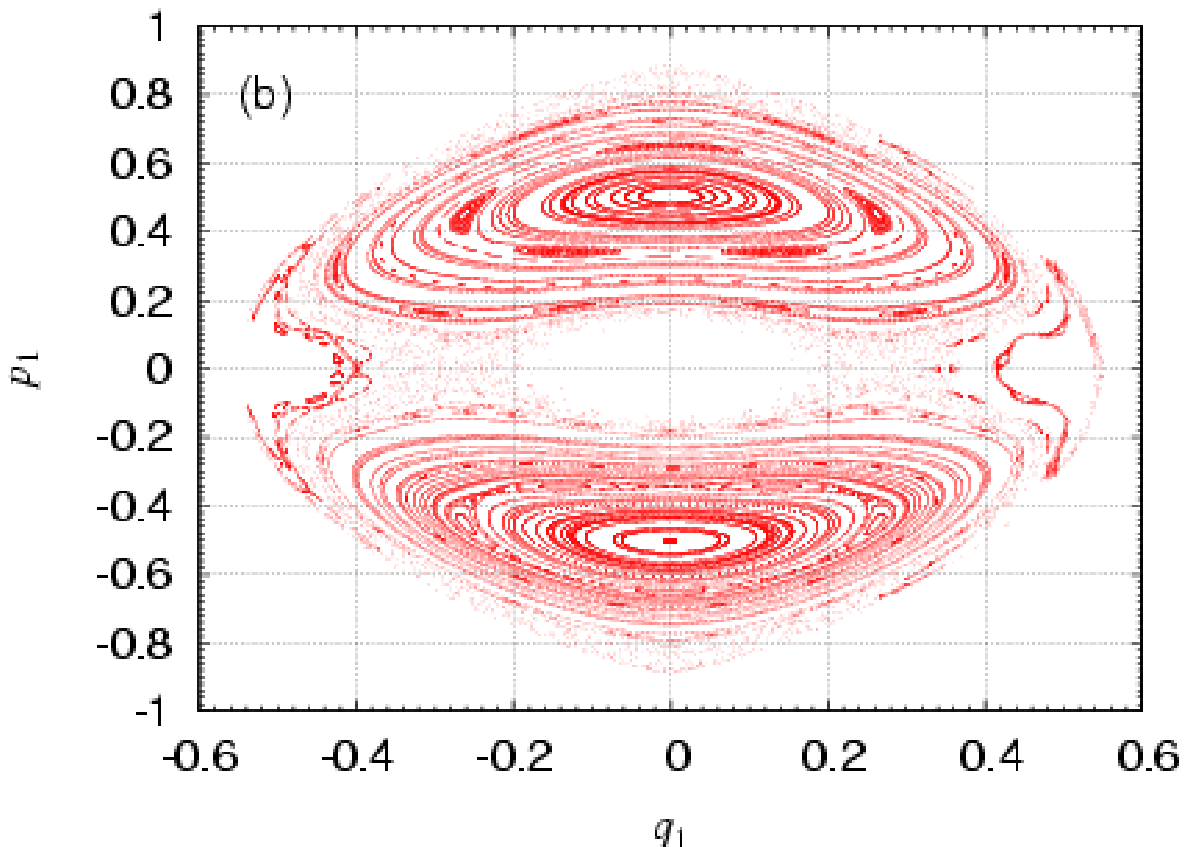}
\end{minipage}
 \hfill
\begin{minipage}{.47\textwidth}
\includegraphics[width=\linewidth]{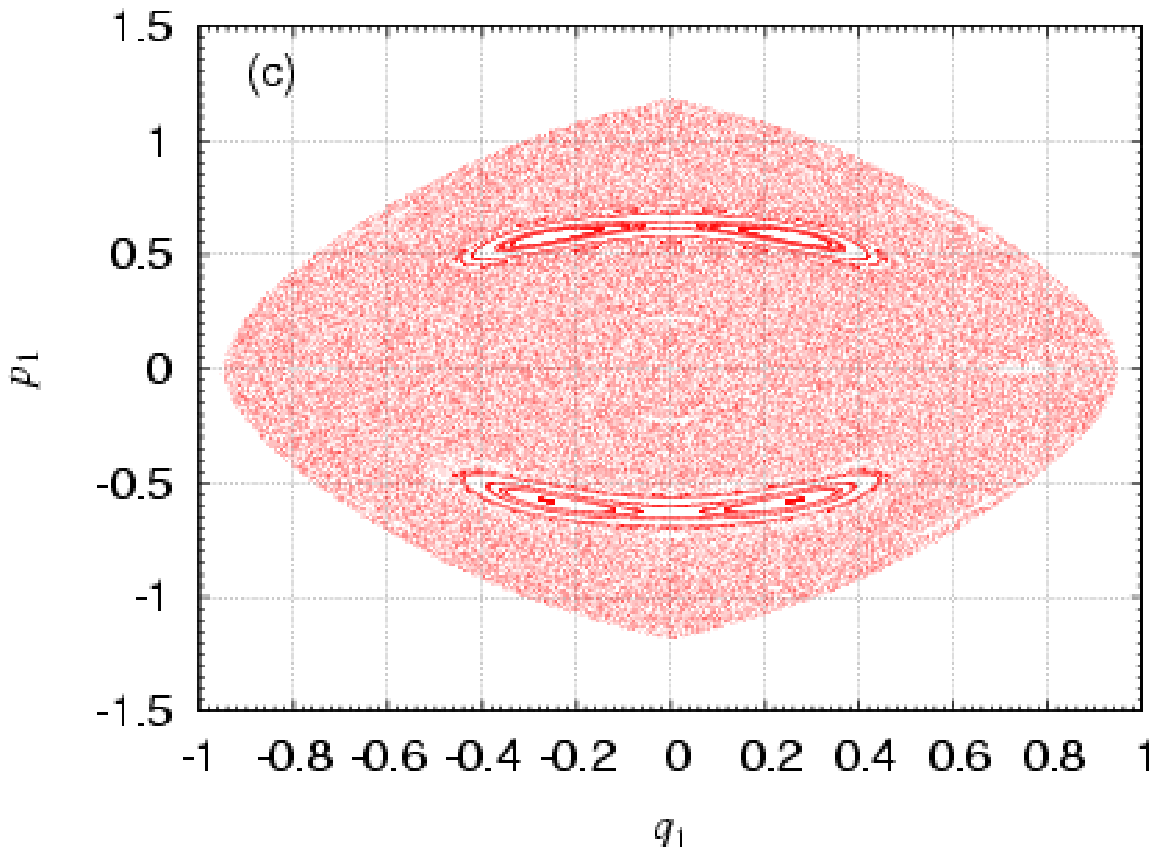}
\end{minipage}
 \caption{Poincar{\'e} sections at $p_2=0.$ $\Omega/\omega=1$: (a)$\epsilon=0.01$,(b)$\epsilon=0.4$,(c)$\epsilon=0.7$.}
   \label{fig2}  
\end{figure*}
Equation (\ref{eqn:7_c}) is nothing but the
Schr{\" o}dinger equation $i{\dot \rho}=[H,\rho]$, from which we find the constant of motion
$\mid {\bf r}\mid=1$. In the study of quasi-classical dynamics, our interest lies
in qualitative comparison between the systems with and without the anharmonic
term, and therefore we confine to the adiabatic limit $\frac{dz}{dt}=0$, that is, ${\bf r}=({\bf r_{\perp}},z_0)$ with
${\bf r_{\perp}}^2=1-{z_0}^2$. Further, 
we find from (\ref{eqn:7_c}),
\[\frac{dz}{dt}{\bf e}_z=k{\bf Q}\times {\bf r_{\perp}}=0\]
which is satisfied only when
${\bf r_{\perp}}\parallel {\bf Q} $. Thus the adiabatic limit is equivalent
to
\begin{equation}
{\bf r_{\perp}}=\sqrt{1-{z_0}^2}\frac{{\bf Q}}{Q},
\label{eqn:8}
\end{equation}
where $Q=|{\bf Q}|$.
Thanks to (\ref{eqn:8}),  (\ref{eqn:7_b})
reduces to
\begin{equation}
\frac{d{\bf P}}{dt}=-\frac{dV({\bf Q})}{d{\bf Q}}
-\tilde{k}\frac{{\bf Q}}{Q}
\label{eqn:9}
\end{equation}
with the renormalized coupling $\tilde{k}=k\sqrt{1-{z_0}^2}$.
Consequently, the classical equation of motion for the present model
can be expressed only by ${\bf Q, P}$ and consists of a set of
(\ref{eqn:7_a}) and (\ref{eqn:9}). This set has the first integral
of motion or the total energy
\begin{equation}
E=\frac{{\bf P^2}}{2}+V({\bf Q})+\tilde{k}\frac{{\bf Q}^2}{Q}.
\label{eqn:10}
\end{equation}

 The following analysis depends on the type of the potential $V({\bf Q})$.
Firstly, we investigate the system with
the harmonic potential only, i.e., $V({\bf Q})=V_0({\bf Q})$. In this case, in addition to the total energy 
(\ref{eqn:10}),
we have another constant of motion, i.e., the orbital angular momentum
\begin{equation}
J_z=({\bf Q}\times{\bf P})_z.
\label{eqn:11}
\end{equation}
The number of constants of motion agrees with the degrees of freedom(two). Therefore the system
is integrable \cite{3,18,19}, showing only regular motions. 

Then, we investigate the system with
the anharmonic potential $V({\bf Q})=
V_0({\bf Q})+V_A({\bf Q})$.
The trigonal field on
the 2-d plane $(Q_1,Q_2)$ is invariant only to operations of the cubic
group\cite{4}. Owing to this breaking of continuous circular symmetry,
the angular momentum $J_z$ in (\ref{eqn:11}) is not a constant of motion,
which makes the system nonintegrable. It should be noted:
the lattice system without
coupling  with the electronic degree of freedom
is identical to the H{\'e}non-Heiles system whose
dynamical features have been intensively studied
in a context of chaos theory\cite{3,10}.

The present system has two control parameters, i.e.,
the coupling constant $\tilde{k}$ between
electronic and vibrational degrees of freedom and
the nonlinearity parameter $b$ responsible for the trigonal field.
However, the simple scaling below
lets them merge to a relevant single parameter.
Let the coordinates
$(Q_1,Q_2)$ be transformed to $(q_1,q_2)$ through
$\sqrt{\frac{b}{\tilde{k}}} Q_1=q_1,\sqrt{\frac{b}{\tilde{k}}} Q_2=q_2$.
The total energy is then written as
\begin{equation}
  E=\frac{1}{2}\frac{\tilde{k}}{b}\left(\frac{d{\bf q}}{dt}\right)^2+
  \frac{\omega^2}{2}\frac{\tilde{k}}{b}{\bf q}^2+\frac{{\tilde{k}}^{3/2}}{\sqrt{b}}
  \left[\frac{1}{2}\sqrt{q_1^2+q_2^2}+\frac{q_1^3}{3}-q_1q_2^2\right].
  \label{eqn:12}
\end{equation}
Next, define the scaled time $\tau=({\tilde{k}}b)^{ 1/4}t$
and the scaled momentum ${\bf p}=\frac{d{\bf q}}{d\tau}$.
Finally, by scaling the energy and the angular frequency as
\begin{subequations}
\begin{align}
\epsilon&=Eb^{1/2}{\tilde{k}}^{-3/2} \label{eqn:13a}\\
\Omega^2&=\omega^2{\tilde{k}}^{-1/2}b^{-1/2}, \label{eqn:13b}
\end{align}
\end{subequations}
the total energy is expressed as
\begin{equation}
  \epsilon=\frac{1}{2}{\bf p}^2+\frac{1}{2}\Omega^2{\bf q}^2+
  \frac{1}{2}\sqrt{q_1^2+q_2^2}+\frac{q_1^3}{3}-q_1q_2^2.
\label{eqn:14}
\end{equation}
Eliminating $\tilde{k}$ between (\ref{eqn:13a}) and (\ref{eqn:13b}),
we find
$\epsilon=\left(\frac{\Omega}{\omega}\right)^6Eb^2$,
which tells that the enhancement of the anharmonic term is equivalent to the
increase of the scaled energy $\epsilon$ under
a fixed value of  $\frac{\Omega}{\omega}$. In our numerics below,
we choose $\frac{\Omega}{\omega}=1$.
Figure \ref{fig2} shows
Poincar{\'e} surface of sections for energies $\epsilon = 0.01, 0.4$ and
$0.7$.
While in Fig.\ref{fig2}(a) we find
almost all trajectories to be regular, Fig.\ref{fig2}(b)
shows that Kolmogorov-Arnold-Moser (KAM) trajectories
begin to collapse and some trajectories become
chaotic. Finally in Fig.\ref{fig2}(c)
chaotic trajectories dominate almost all phase space. These results imply that
the domain of chaos expands gradually with
increasing the scaled energy or the anharmonicity
under a fixed value of the scaled angular frequency.

We have also solved Eq.(7) without use of the adiabatic approximation. Then, the
degree of freedom becomes three. In case that
$k$ and $b$ are less than unity, the relative fraction of chaos in
phase space is quite small: We call this behavior ``partial chaos''.
However, if both $k$ and $b$ are as large as $5$, the global chaos as seen in
Fig.\ref{fig2}(c)
appears. In case of $b=0$, we find no indication of irregularity. We shall now proceed to investigate a quantal counterpart
of these classical features.


\section{Quantum Systems}
In this Section we investigate the quantal manifestation of chaos and of regular
motions without resorting to the adiabatic limit. The construction of the energy matrix is as follows(see \cite{7}
and \cite{25}): The basis wavefunction is
described as the products of electronic wavefunctions $u_{\pm}({\bf r})$ and
vibrational ones $\phi({\bf Q})$. The former is given in terms of $u({\bf r})$
and $v({\bf r})$ below (\ref{eqn:1}) as 
\begin{equation}
 u_{\pm}({\bf r})=\frac{1}{\sqrt{2}}(u({\bf r})\pm iv({\bf r})), \label{eqn:15}
\end{equation}
while the latter is the eigenstate with the eigenvalue $E_{nm}=n\hbar\omega$ for the 2-d harmonic oscillator:
\begin{equation}
 \phi_{n,m}(\rho,\theta)=F_{n|m|}(\rho)e^{im\theta}, \label{eqn:17}
\end{equation}
where $n=1,2,\ldots,$ and $m=n-1,n-3,\ldots,-n+1$. $F_{n|m|}(\rho)$
is the confluent hypergeometric function\cite{25}.
Thus, the basis wavefunctions for the present model are given by
\begin{equation}
 \Phi_{n,m}^{\pm}=u_{\pm}({\bf r})\phi_{n,m}(\rho,\theta).\label{eqn:19}
\end{equation}
Corresponding  to (\ref{eqn:1}), the interaction matrix $H_{J-T}$ is expressed as 
\begin{align}
 H_{J-T}&=V_u({\bf r})Q_1+V_v({\bf r})Q_2 \nonumber \\
 &=\frac{\rho}{\sqrt{2}}\left[V_{u-}({\bf r})e^{i\theta}-V_{u+}({\bf
 r})e^{-i\theta}\right] \label{eqn:20}
\end{align}
with the matrix elements of $V_{u\pm}$ given by 
\begin{align}
 \langle u_{\pm}|V_{u+}({\bf r})|u_{\pm}\rangle&=\langle u_{\pm}|V_{u-}({\bf r})|u_{\pm}\rangle=0 \nonumber \\
 \langle u_{\mp}|V_{u\pm}({\bf r})|u_{\pm}\rangle&=\mp\sqrt{2}k. \label{eqn:21}
\end{align}
\begin{figure}
  \begin{minipage}{.40\textwidth}
   \includegraphics[width=\linewidth]{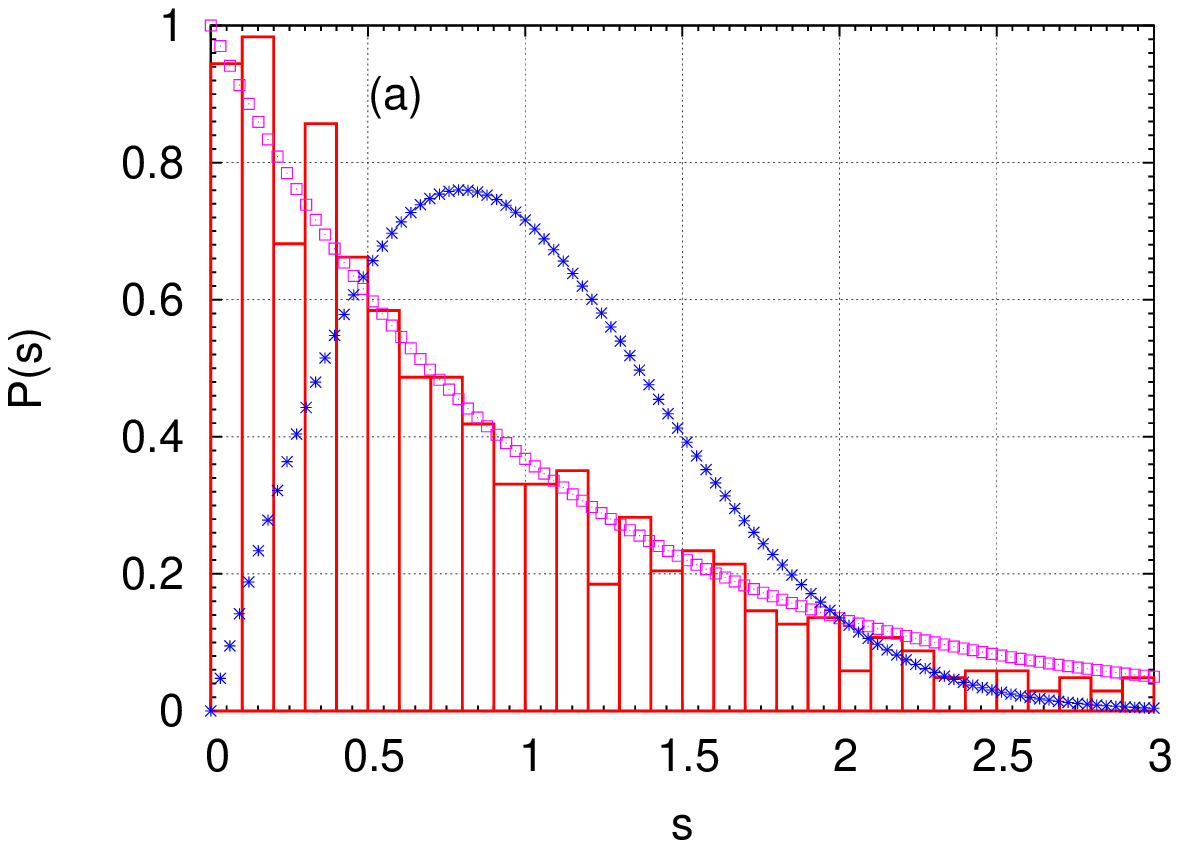}
   \includegraphics[width=\linewidth]{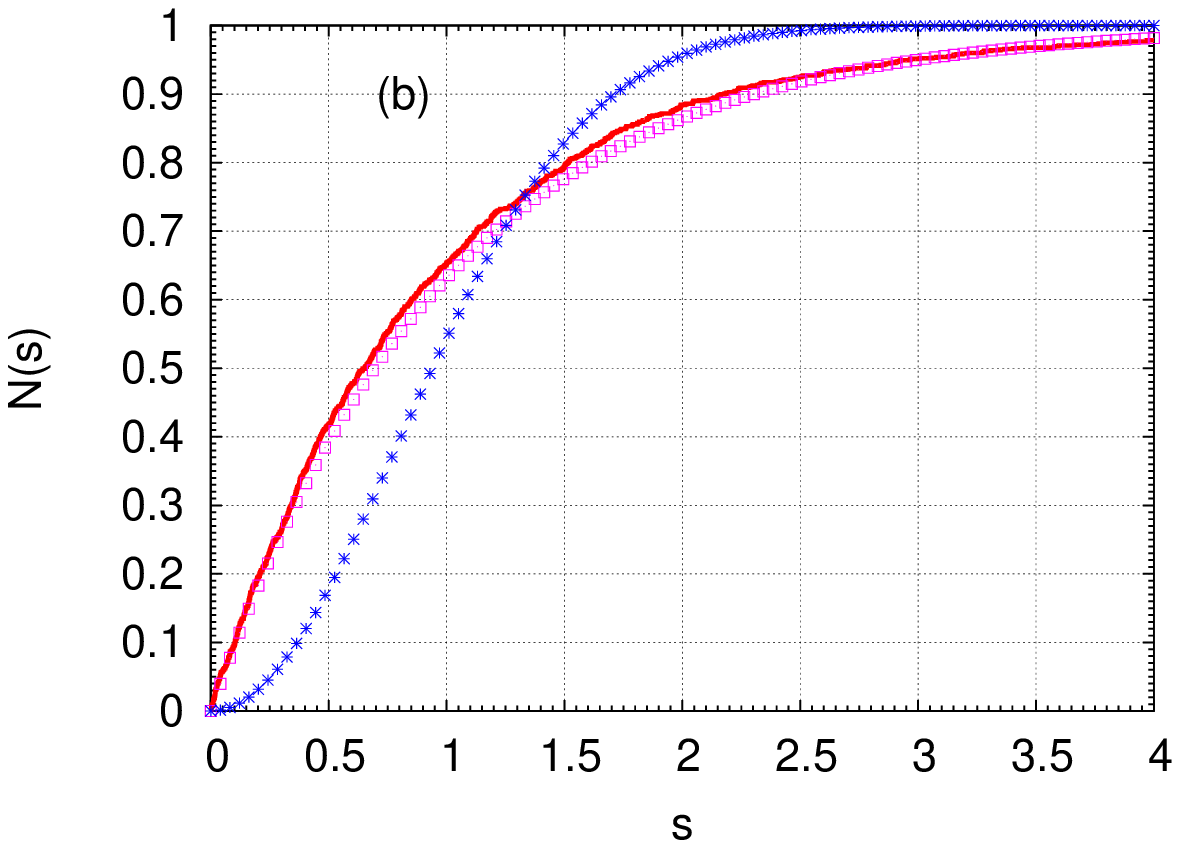}
   \caption{(a) Histograms of level spacing distribution for $k=4,b=0$(without
   trigonal field). Curves of Poisson distribution and GOE are also drawn by
   open rectangles and crosses, respectively. (b) Integrated level spacing
   distribution with use of data in Fig.3(a).}
  \label{fig3}
  \end{minipage}
\end{figure}
As for the lattice-vibration factor $\rho e^{\pm i\theta}$ in (\ref{eqn:20}),
we have
non-vanishing matrix elements as 
\begin{align}
\langle\phi_{n,m}|\rho e^{-i\theta}|\phi_{n+1,m+1}\rangle&=\langle\phi_{n+1,m+1}|\rho
 e^{i\theta}|\phi_{n,m}\rangle \nonumber \\
&=\left[\frac{\hbar}{2\omega}(n+m+1)\right]^{1/2}, \nonumber \\
\langle\phi_{n,m}|\rho e^{-i\theta}|\phi_{n-1,m+1}\rangle&=\langle\phi_{n-1,m+1}|\rho
 e^{i\theta}|\phi_{n,m}\rangle \nonumber \\
&=\left[\frac{\hbar}{2\omega}(n-m-1)\right]^{1/2}. \nonumber 
\end{align}
As a result, the matrix elements for (\ref{eqn:20}) are 
\begin{align}
& \langle\Phi_{n,m}^{+}|H_{J-T}|\Phi_{n',m'}^{-}\rangle=k\langle\phi_{n,m}|\rho
 e^{i\theta}|\phi_{n',m'}\rangle \nonumber \\
&=
 k\{\frac{\hbar}{2\omega}\left[n\pm(m-1)\right]\}^{1/2}\delta_{n',n\mp1}\delta_{m',m-1} \\
& \langle\Phi_{n,m}^{-}|H_{J-T}|\Phi_{n',m'}^{+}\rangle=k\langle\phi_{n,m}|\rho
 e^{-i\theta}|\phi_{n',m'}\rangle \nonumber \\
&=
 k\{\frac{\hbar}{2\omega}\left[n\pm(m+1)\right]\}^{1/2}\delta_{n',n\pm1}\delta_{m',m+1}. \label{eqn:21-2}
\end{align}
If we assign the quantum numbers $j=\pm 1$ to $\Phi_{n,m}^{\pm}$, $H_{J-T}$
without the anharmonic term connects the states with the same quantum number,
$\ell=m-(1/2)j\quad(j=\pm 1)$. As discussed by Longuet-Higgins\cite{7}, the present matrix decomposes
into matrices labeled by quantum number $\ell$. For any given value of
$\ell$, $m$ can take two values, $m=\ell-1/2$ and $\ell+1/2$ corresponding
to $j=-1$ and $+1$, respectively. Thus, the $p$-th eigenfunction for a given
$\ell$ is expressed as $\Psi_{p,\ell}$. If we consider the trigonal field (\ref{eqn:3}),
levels with $\ell \pm 3N(N=1,2,3,\ldots)$ are coupled to levels of $\ell$, as discussed in
\cite{7}. The energy matrix is decomposed into only three
irreducible presentations, $A_g,B_g$ and $E_g$.
By the exact diagonalization of each submatrix for the Hamiltonian
including $H_{J-T}$, we can get eigenvalues and eigenvectors. The result
depends on two parameters, the coupling constant $k$
and the strength of the trigonal field $b$.
\begin{figure}
 \begin{minipage}{.40\textwidth}
   \includegraphics[width=\linewidth]{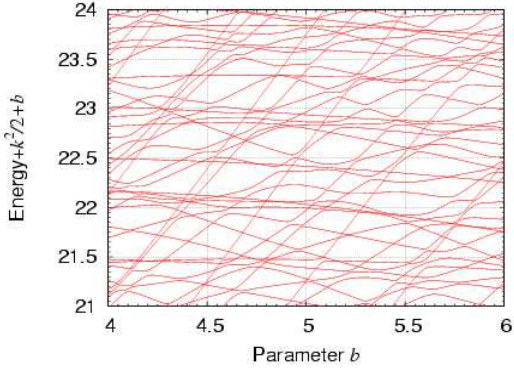}
   \caption{Dependence of eigenvalues($\sim p=120$) on anharmonic parameter $b$.}
   \label{fig4}
  \end{minipage}
\end{figure}
\begin{figure}
  \begin{minipage}{.40\textwidth}
   \includegraphics[width=\linewidth]{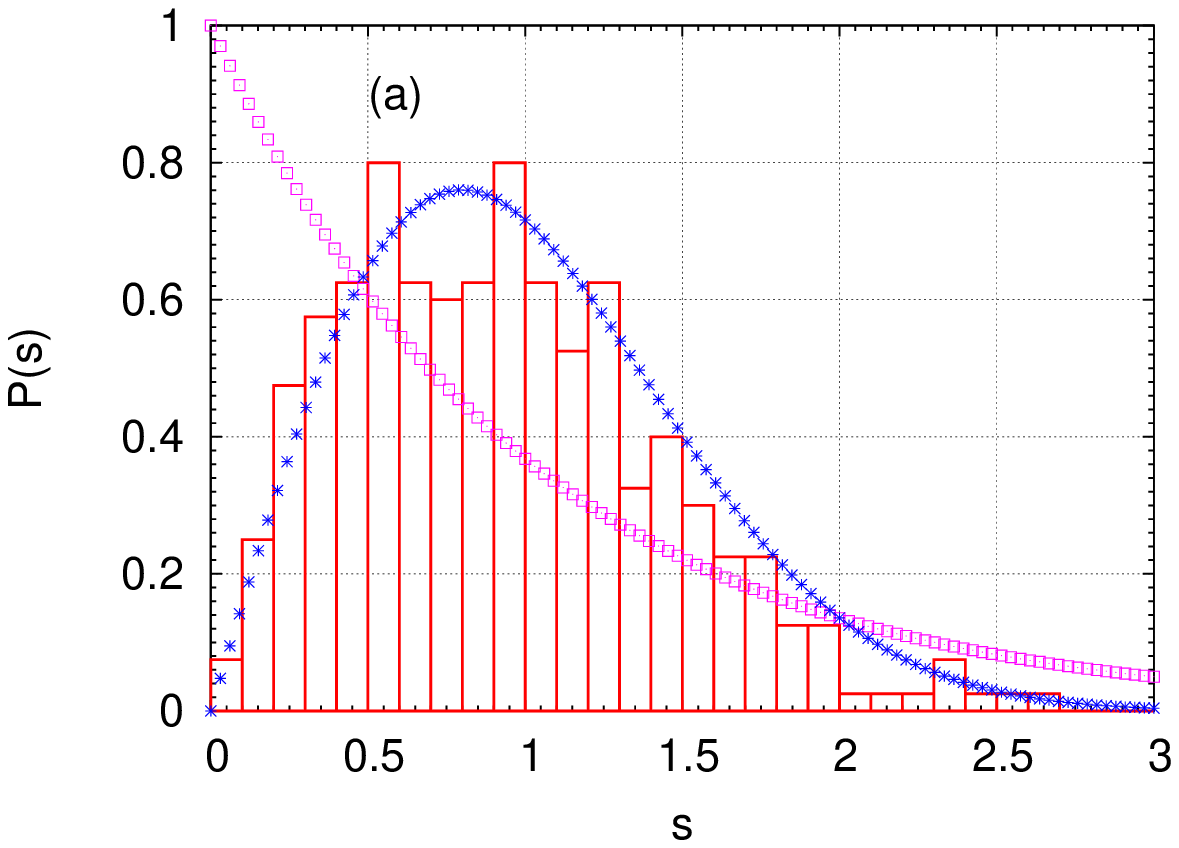} 
   \includegraphics[width=\linewidth]{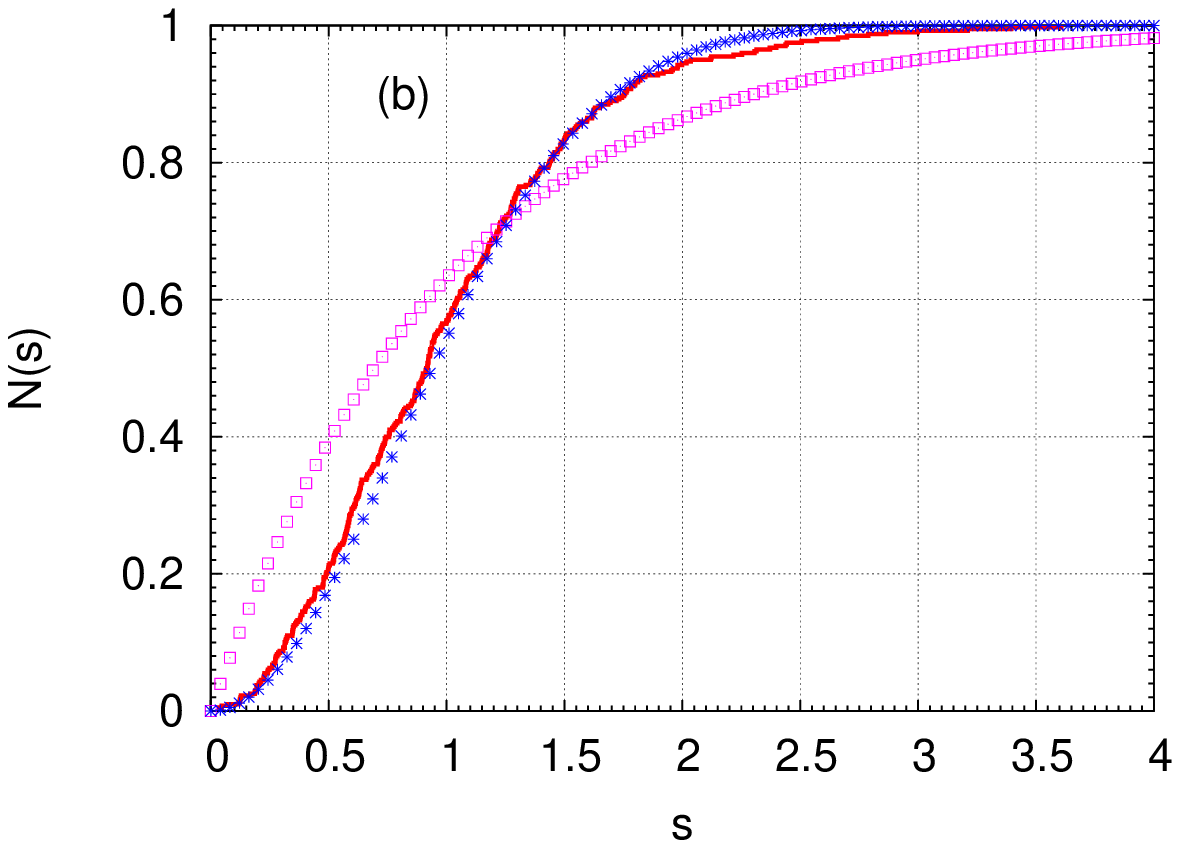}
   \caption{(a) Histograms of level spacing distribution; (b) Integrated level spacing distribution for $k=4,b=5$(with
   trigonal field). Curves of Poisson distribution and Wigner
   distribution are also drawn by open rectangles and crosses, respectively.}
   \label{fig5}
  \end{minipage}
\end{figure}
Here, we concentrate on the nearest-neighbor level spacing
distribution $P(s)$, which plays a prominent role in
the quantum description of classically chaotic quantum systems. The random matrix theory presents a natural framework for describing
fluctuation properties of spectra of quantum systems, whose corresponding
classical model exhibits chaotic behaviors. In fact the correlations in Gaussian ensembles of
random matrices are found to match very closely the empirical correlations
among energy levels in classically chaotic systems. If the phase
space is totally chaotic, the distribution of level spacings is
Wigner-like(GOE),
\begin{equation*}
 P_w(s)=\frac{\pi}{2}s\exp\left(-\frac{\pi}{4}s^2\right).
\end{equation*}
By contrast, in classically regular quantum systems the levels are
independent of each other, and therefore the spacings obey Poisson distribution,
\begin{equation*}
 P_p(s)=\exp(-s). 
\end{equation*}
Considering these standard criteria, we present the level spacing distribution for the
Hamiltonian (\ref{eqn:1}) in the following.

First, we consider the system without trigonal field$(b=0)$.
For the manifold of $\ell=1/2\pm 3m(m=0,1,2\ldots,)$,
Fig.\ref{fig3}(a) shows level spacing distribution characterized by the
Poisson distribution, which tells that this system
is regular in the classical
dynamics\cite{24}. 
Figure \ref{fig3}(b) shows integrated values of histograms in Fig.\ref{fig3}(a), which
well fits the function $1-e^{-s}$.



Then we reveal the role of trigonal field. Figure \ref{fig4} shows the
dependence of eigenvalues on the parameter $b$, where a
multitude of avoid crossings can be found.

Figure \ref{fig5}(a) shows a level spacing distribution in the case of $b=5$.
It should be noted that the distribution perfectly
agree with the Wigner one with $P(s)\to s$ as $s\to 0$. In Fig.\ref{fig5}(b) we
show the corresponding integrated level spacing distribution, which also well fits the Wigner one (GOE). We would
like to point out the following fact:
In the corresponding classical system with $k=4$ and $b=5$ without use of the
adiabatic approximation, the almost whole
regions in the phase space exhibit stochastic behavior. However, the system under consideration is generic: the phase space of the underlying
classical dynamics consists of both regular Kolmogorov-Arnold-Moser(KAM) tori
and chaos. In fact, for smaller values of $b(0<b\lesssim 1)$, we have the Brody
distribution interpolating the Poisson and Wigner limits.

While most of studies on quantum chaos have been
limited to the analyses of level-spacing distributions and of wavefunctions
scars, we shall here embark upon the investigation of experimentally
accessible new indicators. We pay 
attention to
magnetism in the dynamical Jahn-Teller system. In particular we discuss the magnetic $g$-factor in
the next Section\cite{17}.

\section{Electronic orbital angular momentum}
The essential features of the Jahn-Teller coupled
systems are caused by the
interaction between the nuclear motions and
the electrons in non-Kramers doublets for multiplets of $d$-levels.
In particular, the quenching of the electronic orbital angular momentum $-$the
so-called magnetic $g$-factor$-$ in
those doublets is the fundamental subject in magnetism of
transition-metal ionic compounds. As well known, the static
Jahn-Teller effect removes the ground state degeneracy of electronic 
states.
This effect leads to the complete disappearance of the orbital angular
momentum, leaving only spin degree of freedom remains.
On the other hands the dynamic Jahn-Teller systems for the relatively weak
coupling have a possibility of the nonvanishing orbital angular momentum in
non-Kramers doublets because of the continuous distortion of lattices.
In fact, in the absence of a trigonal field the $E_{g}\otimes e_{g}$ system has the freedom
of the continuous distortion of lattices along the continuous minima of the
adiabatic "Mexican hat" potential.

Washimiya\cite{17} payed his attention to the 
dependence of
expectation value of this orbital angular momentum $\langle L_{z}\rangle$ in the excited levels
$\Psi_{p,\ell}$ derived in Section IV.
He pointed out the oscillatory behavior of $\langle L_{z}\rangle$
with increasing energy levels. The non-vanishing 
values $\langle L_{z}\rangle$ are given as
\begin{eqnarray}
\langle L_{z}\rangle_p&=&\langle\Psi_{p,\ell=1/2}|L_z|\Psi_{p,\ell=1/2}\rangle \nonumber \\
  &=&\left[\sum_{n=1}^{\infty}(-1)^na_{n,p}^2\right]\Xi, \label{eqn:24}\\
  &&p=1,2,3,\cdots, \nonumber
\end{eqnarray}
for the vibronic state of $\ell =1/2$.
Here, $a_{n,p}$'s are the coefficients of the harmonic oscillator functions.
As already mentioned, the vibronic wavefunction is given by
\begin{eqnarray}
\Psi_{p,\ell}&=&a_{1p}u_{-}({\bf r})\phi_{1,0}(\rho,\theta)+a_{2p}u_{+}({\bf r})\phi_{2,1}(\rho,\theta) \nonumber \\
&+&a_{3p}u_{-}({\bf r})\phi_{3,0}(\rho,\theta)+a_{4p}u_{+}\phi_{4,1}(\rho,\theta)+\cdots. \label{eqn:25}
\end{eqnarray}
In (\ref{eqn:24}), $\Xi$ is the elements in the principal diagonal of the matrix.
($\Xi=\langle u_+|L_z|u_+\rangle=-\langle u_-|L_z|u_-\rangle$.)
In short, the differences of $|a_{n,p}|^2$ between even- and odd- numbers of
$n$ in the eigenfunction labeled by $p$ play an essential role. As $p$ 
increases, the oscillatory behavior of the angular momentum is found.
The periods of this oscillation are far long in comparison with the variation
by odd and even numbers for the small coupling $k$.
With increasing $k$, the absolute value of the angular momentum decreases.
The origin of this oscillation has not been clarified up to now, though Washimiya's
finding is essential for understanding properties of this vibronic systems.

In what follows, we show a more detailed calculation of this oscillatory 
behavior.
Furthermore, the effect of the trigonal field is discussed in consideration
that this field destroys the continuous circular symmetry and that
the angular momentum $\ell$ is not a good quantum
number. Under that condition, we introduce the density of $g$-factor
$g(\varepsilon)d\varepsilon$ for the electronic orbital angular momentum in the
energy range between $\varepsilon$ and $\varepsilon + d\varepsilon$ as 
\begin{equation}
 g(\varepsilon)d\varepsilon = {\sum_p}'\left|\sum_{n=1}^\infty(-1)^na_{\ell=1/2,n,p}^2\right|d\varepsilon.
\end{equation}
Here, the summation of $p$ is taken over the corresponding energy range. It
should be noted that levels for $\ell'=1/2\pm 3N(N=1,2,\ldots)$ are
mixed with the levels for $\ell=1/2$ in the presence of the trigonal field. 

\begin{figure*}
  \begin{minipage}{.47\textwidth}
   \includegraphics[width=\linewidth]{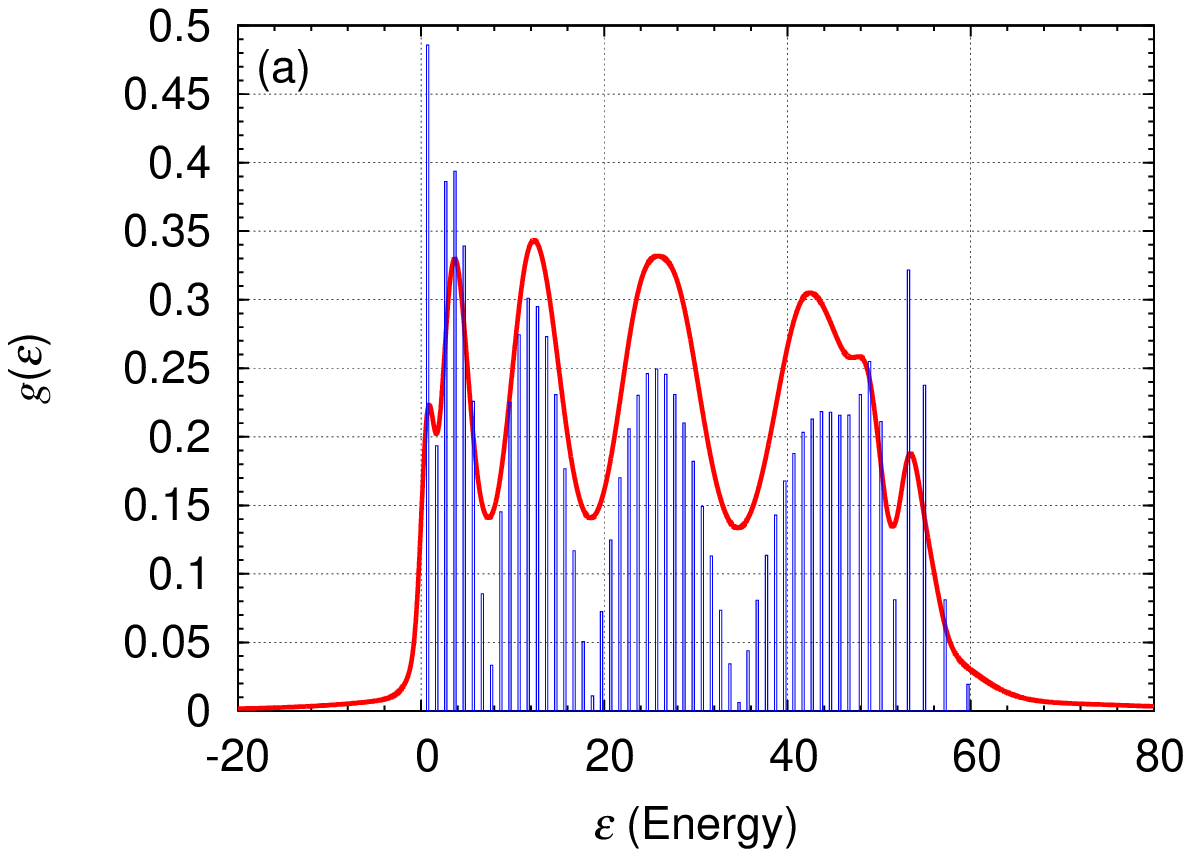}
   \includegraphics[width=\linewidth]{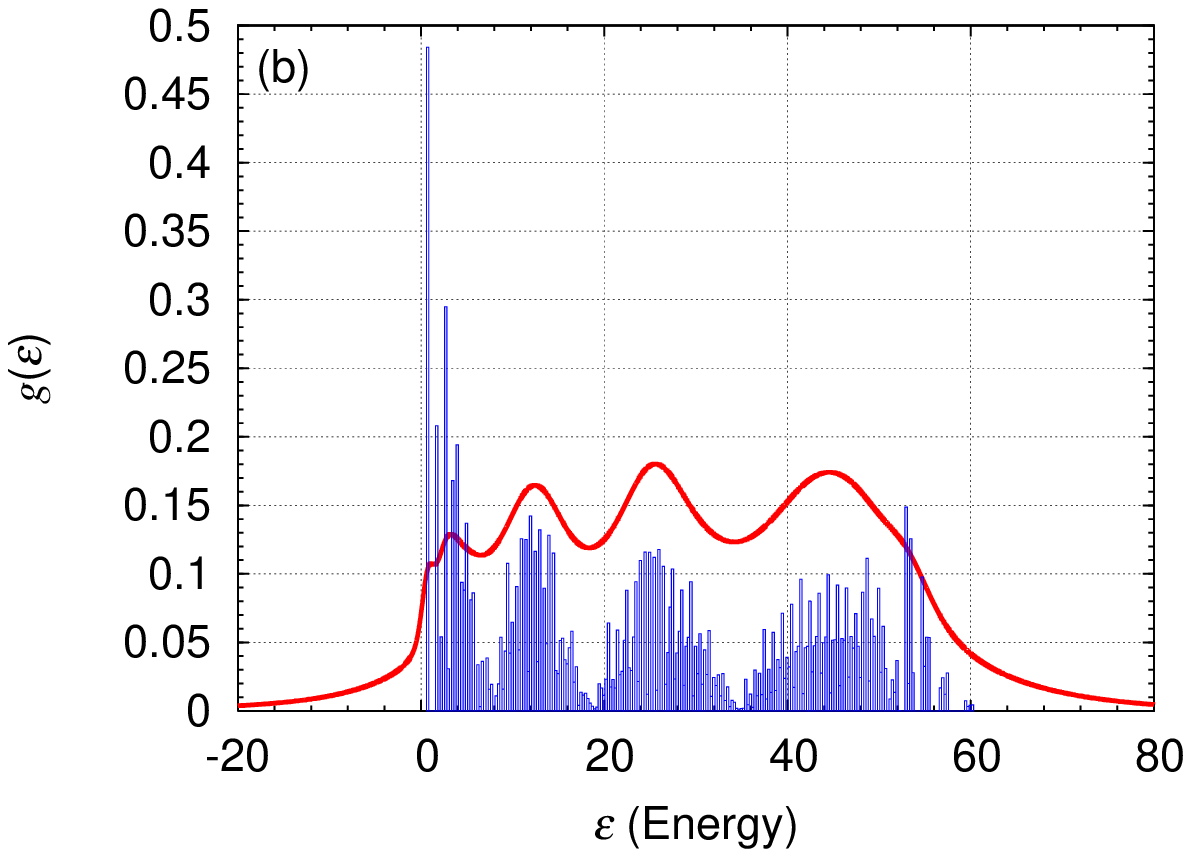}
  \end{minipage}
 \hfill
  \begin{minipage}{.47\textwidth}
   \includegraphics[width=\linewidth]{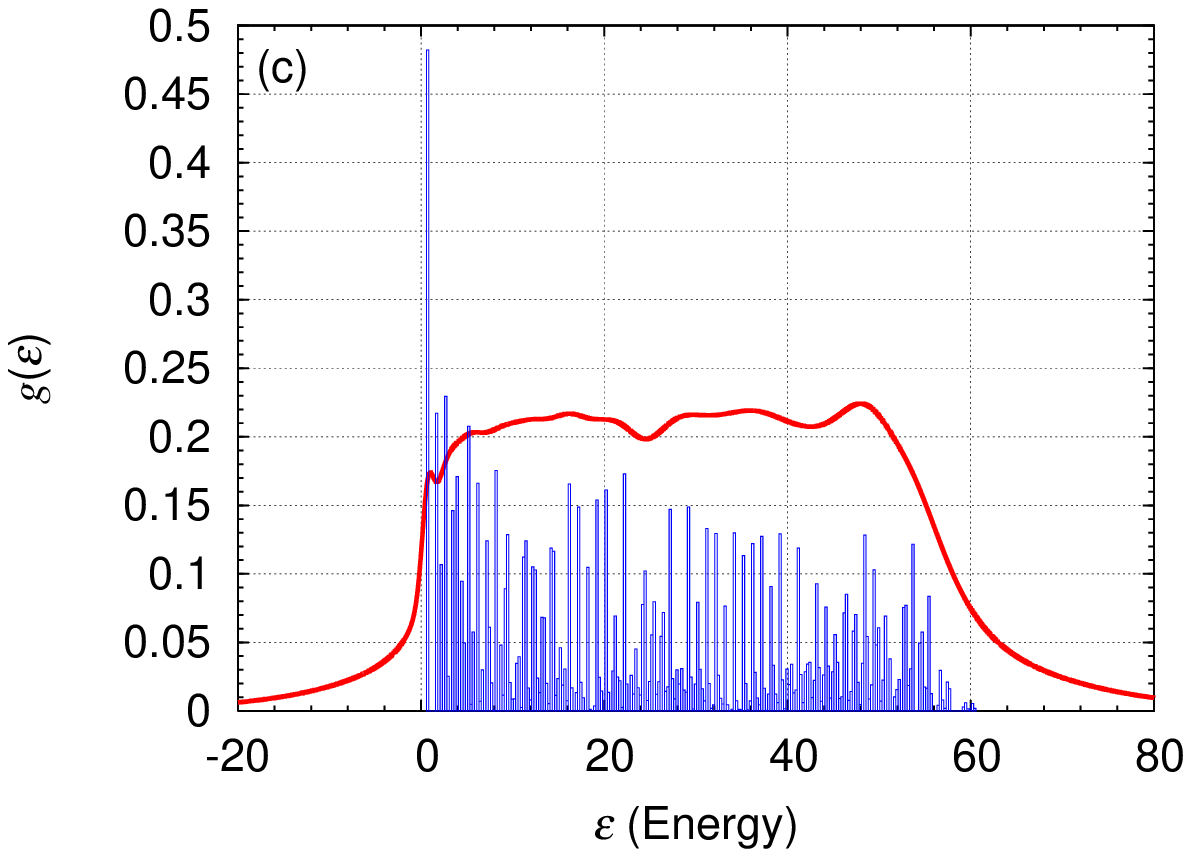}	
   \includegraphics[width=\linewidth]{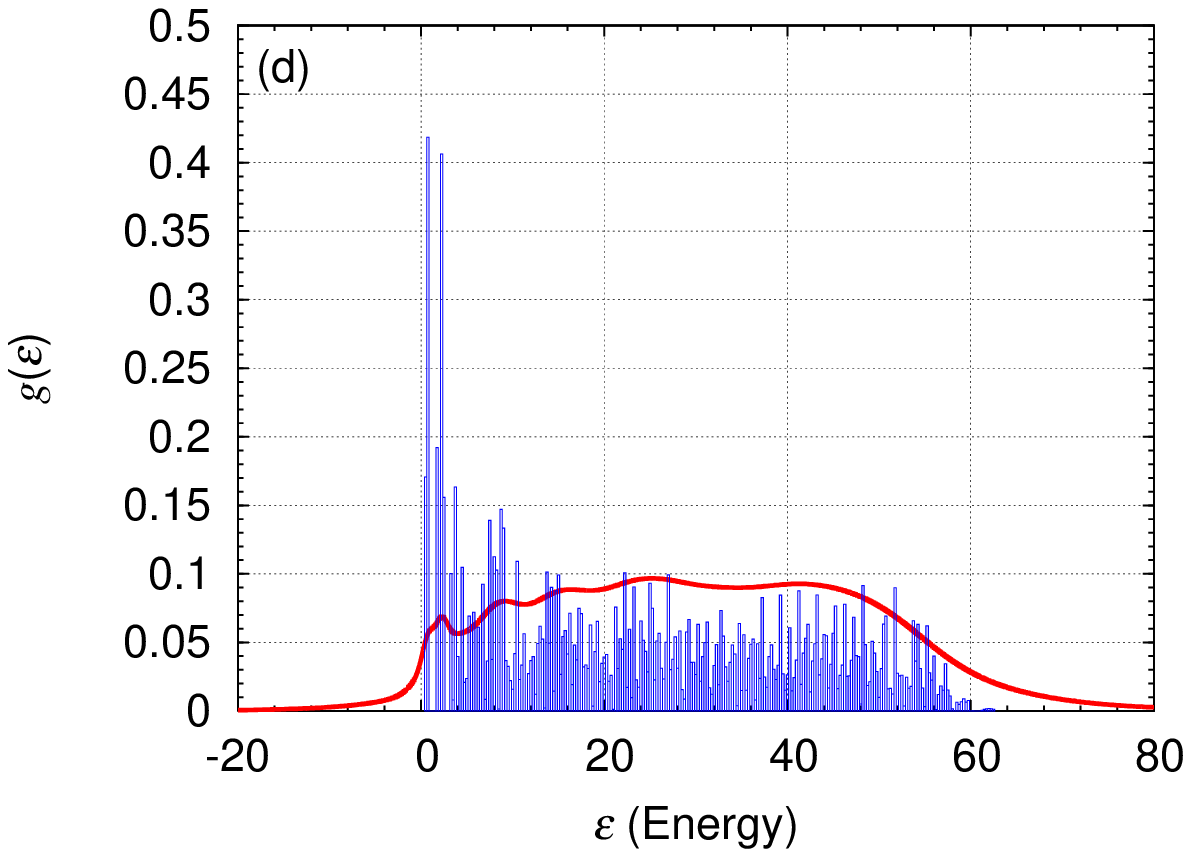}
  \end{minipage}
\caption{$g$-factor $g(\varepsilon)d\varepsilon$($\varepsilon=0.25$) for the
 electronic orbital angular momentum. $\ell=1/2$ and $k=0.707$. (a),(b),(c)
 and (d) correspond to $b=0,0.2,0.3$ and $1.41$, respectively. Envelop-functions composed for Gaussian
 distribution types are also depicted.}
\label{fig7}
\end{figure*}

We show calculated results of $g(\varepsilon)$ at $k=0.707$ in Fig.\ref{fig7}:
In Fig.\ref{fig7}(a), the regular oscillatory behavior of $g(\varepsilon)$
for $b=0$ as a function of level $p$, which was
reported by Washimiya\cite{17} three decades ago, is reproduced. 
In addition to histograms for $g(\varepsilon)$, we also show the envelop-functions
constructed by
Gaussian coarse-graining of each peak. Figures \ref{fig7}(b),(c) and (d)
show the $\varepsilon$ dependence of $g(\varepsilon)$ in the presence of
trigonal field with $b=0.2$, $0.3$ and $1.41$, respectively. Here, we can find the suppression of regular oscillation with
increasing $b$.

The emergence of irregular oscillation and
the suppression of $g$-factor with increasing $b$ reflect the underlying
chaotic behavior in Section III. We find the important fact: while the
level statistics can reach the Wigner distribution for sufficiently-large
value $b$, the suppression of the regular oscillation of $g$-factor easily
occurs for relatively small value $b$. In fact the values of
$b=1.41,k=0.707$ ($\Omega/\omega=1,\quad k \times b=1$) in Fig.\ref{fig7}(d)
are much less than the values $b=5,k=4$ that guarantee the Wigner distribution
in the case without the adiabatic approximation. As a result, the
irregular oscillation of $g$-factor is a precursor of quantum chaos, namely,
the suppression of its regular oscillation occurs even when the classical
phase space accommodates a partial chaos. 
\section{Summary and Discussions}
We have examined the Jahn-Teller $E_g \otimes e_g$ system from a viewpoint of
classical chaos and its quantization.
Both the systems  (A) without an trigonal anharmonic term  and
(B) with it are investigated.

The classical phase space is
strictly regular for (A) and nonintegrable and chaotic for (B).
In general, the system (B) is mixed of regular KAM tori and chaos.
The relative fraction of chaos in the phase space increases as the energy or
the strength of the anharmonicity is increased.
In the adiabatic approximation, the full
chaos can occur easily because of enhancement of non-linear effects due to
the strong constraint. Without such an approximation, we can find the full
chaos for relatively large values of $k$ and $b$, and obtain only the partial
chaos, if we adopt small values to $k$ and $b$.

For the corresponding quantum systems, the level spacing distributions are
shown to be of Poisson and Wigner type for (A) and (B), respectively.
We find that the Wigner distribution is available as well by adopting the relatively large values of $k$ and $b$.
On the other hand, the dependence of $g$-factors on energy
$\varepsilon$ is a quite sensitive indicator of the symptom of chaos
in comparison with the level spacing distribution. In the system (A)
$g$-factor shows regular oscillation with respect to excitation energy. By contrast,
in the
system (B) it shows a quenched irregular oscillation for the relatively small
values of $k$ and $b$. Therefore, we propose the magnetic $g$-factor as a new
precursor of quantum chaos, superior to the level-spacing distribution.
We hope these predictions will be verified in future
such as in the experiment of magnetic circular dichromism.

It will be quite important that the precursor of chaos shows up in 
the observable orbital angular momentum in the dynamical Jahn-Teller systems for
transition-metal ionic-compounds. There are other interesting
themes in this system, such as an 
effect of
the chaos on a spectral feature of phonon side bands,
which will also be examined in due course.
\bibliography{qcjt}

\end{document}